\documentclass[fleqn,usenatbib]{mnras}

\usepackage{newtxtext,newtxmath}

\usepackage[T1]{fontenc}
\usepackage{ae,aecompl}

\usepackage{graphicx}	
\usepackage{amsmath}	
\usepackage{amssymb}	
\usepackage{times}

\newcommand{\kms}        {\,{\rm km}\,\,{\rm s}^{-1}}
\newcommand{\gaia}		{\emph{Gaia}}

\title[Galactic escape speed]{The local high velocity tail and the Galactic escape speed}

\author[A. J. Deason et al.]{
Alis J. Deason$^{1}$\thanks{E-mail: alis.j.deason@durham.ac.uk},
Azadeh Fattahi$^{1}$,
Vasily Belokurov$^{2}$,
N. Wyn Evans$^{2}$, \newauthor
Robert J. J. Grand$^{3}$,
Federico Marinacci$^{4}$,
R{\"u}diger Pakmor$^{3}$
\\
$^{1}$Institute for Computational Cosmology, Department of Physics, University of Durham, South Road, Durham DH1 3LE, UK\\
$^{2}$Institute of Astronomy, Madingley Road, Cambridge CB3 0HA\\
$^{3}$Max-Planck-Institut f{\"u}r Astrophysik, Karl-Schwarzschild-Str. 1, D-85748, Garching, Germany\\
$^{4}$ Harvard-Smithsonian Center for Astrophysics, 60 Garden St., Cambridge, MA 02138, USA 
}

\date{Accepted XXX. Received YYY; in original form ZZZ}

\pubyear{2018}

\begin{document}
\label{firstpage}
\pagerange{\pageref{firstpage}--\pageref{lastpage}}
\maketitle

\begin{abstract}
We model the fastest moving ($v_{\rm tot} > 300 \kms$) local ($D \lesssim 3$ kpc) halo stars using cosmological simulations and 6-dimensional \gaia\ data. Our approach is to use our knowledge of the assembly history and phase-space distribution of halo stars to constrain the form of the high velocity tail of the stellar halo. Using simple analytical models and cosmological simulations, we find that the shape of the high velocity tail is strongly dependent on the velocity anisotropy and number density profile of the halo stars --- highly eccentric orbits and/or shallow density profiles have more extended high velocity tails. The halo stars in the solar vicinity are known to have a strongly radial velocity anisotropy, and it has recently been shown the origin of these highly eccentric orbits is the early accretion of a massive ($M_{\rm star}\sim 10^9 M_\odot$) dwarf satellite. We use this knowledge to construct a prior on the shape of the high velocity tail. Moreover, we use the simulations to define an appropriate outer boundary of $2r_{200}$, beyond which stars can escape.  After applying our methodology to the \gaia\ data, we find a local ($r_0=8.3$ kpc) escape speed of $v_{\rm esc}(r_0) = 528^{+24}_{-25} \kms$. We use our measurement of the escape velocity to estimate the total Milky Way mass, and dark halo concentration: $M_{200, \rm tot} = 1.00^{+0.31}_{-0.24} \times 10^{12}M_\odot$, $c_{200}=10.9^{+4.4}_{-3.3}$. Our estimated mass agrees with recent results in the literature that seem to be converging on a Milky Way mass of $M_{200, \rm tot} \sim 10^{12}M_\odot$.

\end{abstract}

\begin{keywords}
Galaxy: fundamental parameters -- Galaxy: kinematics and dynamics
\end{keywords}

\section{Introduction}
Stars with extreme velocities have often been studied in the Milky Way. Akin to our fascination with the most distant, most massive, most luminous --- insert blank --- astronomers are keen to find the fastest stars in the Galaxy \citep[e.g.][]{hattori18, marchetti18, shen18}. However, this pursuit is more than just a record breaking exercise. The fastest moving stars can be related to exotic mechanisms, such as dynamical interactions with the central super massive black hole \citep[e.g.][]{hills88, yu03, brown05}, dynamical interactions between massive stars \citep[e.g.][]{poveda67, leonard90} supernova explosions in binary systems \citep[e.g.][]{blaauw61, portegies00} and even ejection from the Large Magellanic Cloud \citep[e.g.][]{boubert16}. While these mechanisms often produce stars that are unbound from the Galaxy, the fastest ``garden variety" stars are the most prevalent: namely, the high velocity tail of the stellar halo.

The extreme halo stars are bound to the Galaxy, but represent the lowest energy orbits that are capable of reaching the largest extents in the Milky Way. It is for this reason that this population has garnered so much attention: the fastest halo stars in the local vicinity can probe the potential out to the virial radius of the Galaxy.  Indeed, the high velocity stars in the solar neighbourhood present 
one of the only local measures of the gravitational potential at large radii. Historical measurements of the local escape velocity date back to the early 1980s, in the period where the existence of massive dark matter haloes was gaining traction in the astronomy community \citep[e.g.][]{faber79, rubin80}. These early works generally estimated a lower limit on the escape speed by identifying the highest velocity stars in the solar neighbourhood \citep{caldwell81, alexander82, sandage87, carney88}. The seminal work by \citet[][hereafter LT90]{lt90} extended this formalism to produce statistical models for the distribution of stars near the escape speed; this advancement was needed to properly model limited sample sizes that may not include stars that reach the escape velocity, and/or could include spurious measurements due to observational errors. LT90 apply their formalism to $N \sim 30$ high velocity stars with accurate radial velocity measurements and inferred a local escape velocity in the range 450-650 $\kms$.

Two decades later, works by \cite{smith07} and \cite{piffl14} applied the LT90 method to the RAdial Velocity Experiment (RAVE) survey data, finding a local escape speed in the range $\sim 500-600 \kms$. These later works used cosmological simulations to help model the high velocity tail of their stellar halo sample. A similar approach was used in \cite{williams17} to constrain the escape velocity over a wider radial range using Sloan Digital Sky Survey data. In agreement with \cite{smith07} and \cite{piffl14}, they find a local escape velocity of $\sim 520 \kms$. Most recently, \cite{monari18} exploited the new 6-dimensional data from the \gaia\ mission \citep{gaia16, gaia18} to constrain the local escape speed to be $v_{\rm esc}(r_0) = 580 \pm 63 \kms$, where $r_0=8.3$ kpc. \cite{monari18} use the same methodology as \cite{piffl14}, but find a larger escape speed, suggesting that the previous constraints from line-of-velocities only may have underestimated the escape speed (albeit the uncertainties are large).

The above analyses suffer from several potential systematic limitations. First, it is not guaranteed that the tail of the velocity distribution is occupied all the way to the escape velocity. Thus, if there is any truncation in the stellar velocities, the escape speed will be underestimated. Second, although it is only the high velocity tail of the escape speed that needs to be modeled, the stellar distribution need not be smooth and relaxed. Indeed, the presence of substructure in the high velocity tail could significantly bias the results. Third, the estimates are very sensitive to the fastest stars in the sample, so the presence of interlopers (such as unbound stars) or statistical outliers in the data could also effect the derived escape velocity. Despite these apparent shortcomings, there is also warrant for significant optimism. The latest \gaia\ data has revealed that the inner stellar halo is dominated by the material from one massive ($M_{\rm star}\sim 10^9M_\odot$) dwarf galaxy accreted 8-10 Gyr ago \citep{belokurov18, deason18, haywood18, helmi18}. Thus, there is reason to believe that the stellar material, or at least the majority of it, is well phase-mixed. In addition, the highly eccentric orbits of the stars associated with this massive dwarf are more likely (i.e. relative to more circular orbits) to traverse significant distances in the Galaxy, and can potentially probe out to the very outskirts of the Milky Way. With this in mind, the focus of this contribution is to re-formulate the LT90 analysis using these new observational advancements.

The escape velocity provides a direct measure of the Galactic potential, and hence a common goal of constraining this fundamental parameter is to provide an estimate of the total Milky Way mass. Despite decades of study, the mass of the Milky Way has remained a contentious issue in the literature (see \citealt{bland16} Section 6.3 for a recent review), with quoted mass estimates varying by a factor of $2-3$. Recent progress since the second \gaia\ data release has perhaps relieved some of this tension, with estimates generally ranging from $1-1.5 \times 10^{12}M_\odot$ \citep[e.g.][]{eadie18, malhan18, watkins18, callingham19, posti19, vasiliev19}. However, the significance of this parameter warrants that our community strives to pin down the mass with much greater precision and accuracy. Indeed, the total Milky Way mass is essential to place our Galaxy in context with the general galaxy population, and, moreover, the halo mass is central to our understanding of the $\Lambda$CDM paradigm \citep[e.g.][]{purcell12, wang12}.

In this study, we use a combination of analytical models, cosmological simulations and \gaia\ data to model the high velocity tail of the local stellar halo. Through our analysis we provide a new estimate of the local escape velocity, and, by extension, the total Milky Way mass. The paper is arranged as follows. Section \ref{sec:theory} provides the theoretical background to the form of the high velocity tail, and introduces the LT90 formalism. In Section \ref{sec:sims}, we explore the high velocity tails of accreted stars in the Auriga simulations. We use the simulations to place a prior on the form of the high velocity tail, which is appropriate for the Milky Way. We apply our formalism to \gaia\ data release 2 in Section \ref{sec:gaia} , and provide a new estimate of the local escape speed. In Section \ref{sec:mass}, we relate the escape speed to the total Milky Way mass. Finally, in Section \ref{sec:conc} we summarise the main findings of our work.

\section{Theoretical Background}
\label{sec:theory}

In this work, we use simple models for the velocity distribution of stars near the escape speed. This formalism was first presented in \cite{lt90} (hereafter, LT90), and later extended and adapted by \cite{smith07} and \cite{piffl14}. Here, we provide a brief recap of the LT90 method, and provide some analytical insight into the form of the high velocity tails.

\subsection{Leonard \& Tremaine approximation}
\label{sec:lt90}
LT90 proposed a distribution of space velocities appropriate for a sample of high-velocity stars near the Sun:
\begin{equation}
\label{eq:lt90}
f(v | v_e, k) \propto (v-v_e)^k
\end{equation}
for $v < v_e$. Here, $v$ is the total velocity and $v_e$ is the escape velocity. This form only needs to be valid near $v \sim v_e$, and, when $f$ is a power-law of energy, eqn. \ref{eq:lt90} can be thought of as the first term in a Taylor expansion of $f$ near $v_e$. Here, $k$ is a free parameter, and, as we will show in this work, it is strongly dependent on the form of the underlying distribution function.  Note that \cite{smith07} use a slightly different distribution function, namely $f(v | v_e, k) \propto (v_e^2-v^2)^k$, however we choose to adopt the original LT90 formalism as this provides a better description of the high velocity tails in the simulated haloes (see also \citealt{piffl14} Section 3). The LT90 formalism assumes that the stellar system is described by an Ergodic distribution function, and is thus well-mixed in phase-space. Moreover, this approach assumes that the stellar velocities extend all the way to $v_{e}$. Clearly, these assumptions are not necessarily true, and in the following Section(s) we will discuss these potential limitations in light of recent observations of the Milky Way halo, and in the context of cosmological simulations.

\begin{figure*}
        \centering
        \includegraphics[width=17cm,angle=0]{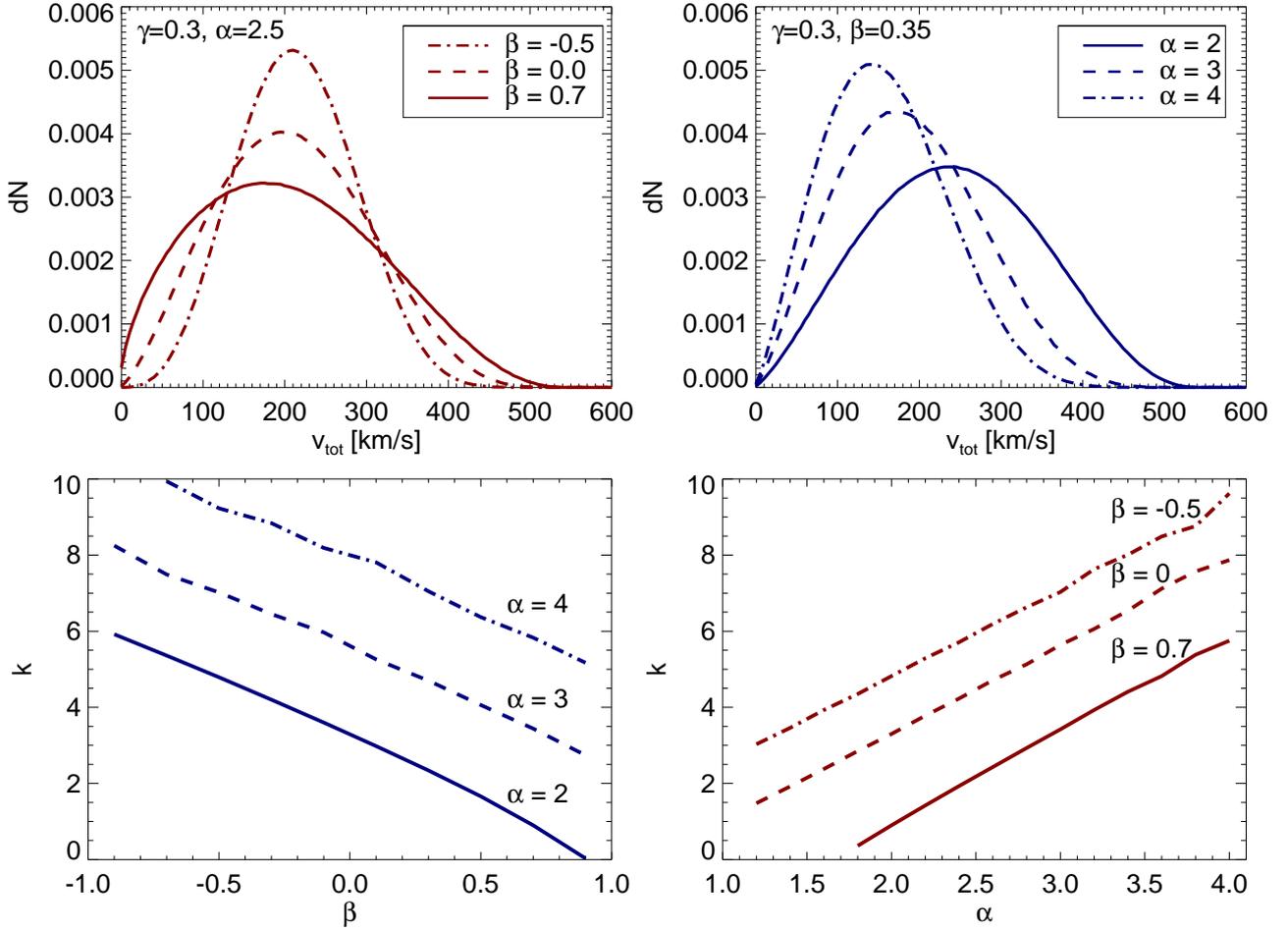}
        \caption[]{Velocity distribution functions (dfs) from the spherical power-law models presented in \cite{evans97}. Here, the dfs are a function of tracer density slope ($\alpha$), velocity anisotropy ($\beta$) and gravitational potential slope ($\gamma$) --- see Eqns. \ref{eq:df1} and \ref{eq:df2}  in the text. We show models with a fixed potential, where $\gamma=0.3$ and $v_{\rm esc} = 550 \kms$. Note that this $\gamma$ is the median value we find in the Auriga simulations at $r \sim 8$ kpc. In the top panels we show the total velocity distribution for fixed $\alpha$ (left) and $\beta$ (right).  The bottom panels show a power-law fit to the high velocity tail (with $v_{\rm tot} > 300 \kms$), of the form $\propto (v_{\rm esc}-v)^k$, for various $\beta$ and $\alpha$.  The systems with highly radial anisotropy and/or shallower tracer density profiles have more extended velocity tails, and thus lower values of $k$ (see text for details).}
          \label{fig:df}
\end{figure*}
\subsection{Maximum likelihood analysis}
In order to constrain $v_e$ and $k$ from a local sample of stars we employ a maximum likelihood method:
\begin{equation}
\mathcal{L} = \prod_{i=1}^{N} f(v_i | v_e, k)
\end{equation}
In practice, we use Bayes' theorem to to derive the probability distributions of the model parameters:
\begin{equation}
P(v_e,k | v_{i=1,...,N}) = \frac{P(v_e)P(k)\prod_{i=1}^{N} f(v_i | v_e,k)}{\int\int P(v_e)P(k)\prod_{i=1}^{N} f(v_i | v_e,k)~\mathrm{d}v_e \mathrm{d} k}
\end{equation}
In the following Section, we introduce an optimal prior $P(k)$ based on cosmological simulations. This approach was also taken by \cite{smith07} and \cite{piffl14}. However, in this work we make use of recent breakthroughs in our understanding of the local halo velocity distribution to form a prior tailored towards our own Galaxy. We find, like previous authors,  that a prior  on $k$ is essential, especially when faced with small number statistics and/or significant velocity errors. Finally, like LT90, we adopt a (weak) prior on $v_e$, $P(v_e) \propto 1/v_e$, which is appropriate for a variable that ranges from $0$ to $\infty$ \citep{kendall77}.

Eqn \ref{eq:lt90} is only valid near $v_e$, so our analysis is performed on stars with $v > v_{\rm min}$. Following \cite{kochanek96} and \cite{smith07} we adopt $v_{\rm min} = 300$ km s$^{-1}$; this cut is chosen to minimize contamination from disc stars, and restrict ourselves to stars close to $v_e$. However, we note that adopting a slightly lower threshold, $v_{\rm min} = 250$ km s$^{-1}$ (cf. \citealt{monari18}), does not significantly affect our results.
\subsubsection{Radial dependence of escape velocity}
In a small enough volume the escape velocity $v_e$ is approximately constant, but more generally $v_e$ is radially dependent, where $v_e=v_e(r) \propto \sqrt{2\Phi(r)}$. In this work, we parametrise $v_e$ as:
 \begin{equation}
 v_e = v_{e,0} \left(r/r_0\right)^{-\gamma/2}
 \end{equation}
where, $r_0=8.3$ kpc is the solar radius, and $v_{e,0}$ is the escape speed at the position of the Sun. Our parametrisation is motivated by the approximate power-law form of the gravitational potential over a small radial range, where $\Phi \propto r^{-\gamma}$. Note that this power-law dependence of the escape velocity was also used by \cite{williams17} over a much larger radial range.

\subsection{Analytical example: spherical, power-law distribution functions}
\label{sec:dfs}
To provide some theoretical insight into the LT90 formalism, we explore the high velocity tails in simple, power-law distribution functions. We adopt the distribution functions introduced in \cite{evans97}, and later adopted in \cite{deason11a}. This model assumes spherical power-laws for the gravitational potential $\left(\Phi(r) \propto r^{-\gamma}\right)$ and tracer density profile $\left(\rho(r) \propto r^{-\alpha}\right)$, and has constant velocity anisotropy $\left(\beta=1-\left[\langle v^2_\phi\rangle + \langle v^2_\theta\rangle\right]/2\langle v^2_r\rangle \right)$. The velocity distribution is given in terms of the binding energy $\left(E = \Phi(r)-0.5v^2_{\rm tot}\right)$ and the total angular momentum $\left(L=\sqrt{L^2_x+L^2_y+L^2_z}\right)$:
\begin{equation}
\label{eq:df1}
F(E,L) \propto L^{-2\beta} f(E),
\end{equation}
where
\begin{equation}
\label{eq:df2}
f(E) = E^{\beta(\gamma-2)/\gamma+\alpha/\gamma-1.5}.
\end{equation}
In the top panels of Fig. \ref{fig:df} we show the total velocity distributions derived from these models. Here, we fix the potential with $v_{\rm esc} = 550$ km s$^{-1}$ and $\gamma=0.3$, and vary $\alpha$ and $\beta$. Note, for illustration, we evaluate this model at a fixed radius, $r=r_0=8.3$ kpc. In the top-left panel we fix $\alpha$ and vary $\beta$, and in the top-right panel we fix $\beta$ and vary $\alpha$. It is clear that the velocity distributions differ when we vary the tracer density profile and/or velocity anisotropy. In particular, although the models all have the same potential (and escape velocity) the forms of the high velocity tails vary significantly.

To explore this further we fix $v_{\rm esc}$ and fit the slope of the high velocity tail ($k$) for each model using Eqn. \ref{eq:lt90}. Here, we use a minimum velocity threshold, $v > 300$ km s$^{-1}$. The bottom panels of Fig. \ref{fig:df} show how $k$ varies with different values of $\alpha$ and $\beta$. Radially anisotropic orbits (higher $\beta$) and/or shallow tracer density profiles (lower $\alpha$) lead to lower values of $k$. The high velocity tails are more populated by stars on highly eccentric orbits (larger $\beta$) because these are biased towards lower energy, and hence larger speeds. This also makes sense physically, as stars on radial orbits can reach to larger distances on their orbits, and have more chance of ``escape". Note that the $L^{-2\beta} \propto v^{-2\beta}$ term in Eqn. \ref{eq:df1} leads to the low velocity form of the velocity distribution, whereby systems with large $\beta$ values also populate the low velocity regime. The net result is a broader distribution for radially anisotropic orbits,  with a strong tail to high velocities. In contrast, the distribution for tangential orbits is more strongly peaked, and does not populate the high velocity (low energy) or low velocity (low angular momentum) regimes.
 In a given gravitational potential, and at fixed $\beta$, more extended tracer populations (smaller $\alpha$) are biased towards lower energies, and hence larger speeds. Thus, when $\alpha$ is low there are more stars that populate the high velocity tail, and $k$ is lower. Again, physically one can imagine that stars drawn from a shallower radial number density distribution are more likely to extend to larger distances (and hence lower energies) on their orbits. 
 
The $k$ values predicted by these spherical, power-law models can be compared to the predictions for a system undergoing violent relaxation. In this case, \cite{jaffe87} and  \cite{tremaine87} show that $k =1.5$. Indeed, \cite{lt90} and \cite{kochanek96} adopt $k$ values that bracket the violent relaxation prediction with $k \in [0.5,2.5]$. These predictions for $k$ were based on self-gravitating systems, rather than the tracer populations considered here. However, this historical range of $k$ agrees with the high $\beta$, low $\alpha$ regime of the power-law dfs shown in Fig. \ref{fig:df}.

Although these models are idealised, they give us an important insight into the high velocity tails of stellar systems. In particular, we see that the power-law slope of the velocities near $v_e$ depends on the velocity anisotropy and density profile of the tracer stars. In our own Milky Way we now have a good handle on these properties, particularly for stars close to the Sun. In the inner regions ($r < 20$ kpc) of the halo the density profile is an approximate power-law with index $\alpha \sim 2.5$ \citep[e.g.][]{deason11b, sesar11, faccioli14, pila15}. We also know that the orbits of local halo stars are highly eccentric \citep[$\beta =0.7$][]{smith09, bond10}. Indeed, recent works using the latest \gaia\ data releases have shown that the stellar orbits in the inner regions of the halo are strongly radial, and the stars in these inner regions are mainly contributed by one massive dwarf progenitor \citep{belokurov18, helmi18}. Thus, importantly, the aforementioned observations can limit the range of $k$ applicable to our own Galaxy. In the following Section, we explore the relation between $k$ and the stellar halo properties further, using the more realistic distributions present in cosmological simulations.

\section{Cosmological Simulations}
\label{sec:sims}

\subsection{Auriga simulation suite}
We use the Auriga simulation suite to explore the high velocity tails of stellar haloes. Auriga is a suite of high resolution Milky Way-mass haloes, spanning a mass range $1 \times 10^{12} <M_{200}/M_\odot < 2\times 10^{12}$. Here, we give a brief description of the simulations and defer the interested reader to \cite{grand17} for more details.

The Auriga suite comprise of $N=30$ re-simulated haloes, which were chosen from the $100^3$ Mpc$^3$ dark matter only periodic box from the EAGLE project \citep{crain15, schaye15}. The candidate haloes were chosen to have a similar mass to the Milky Way, and be relatively isolated at $z=0$: i.e. with no massive objects (greater than half of the parent halo's mass) closer than 1.37 Mpc. The cosmological parameters in the simulation are consistent with the \cite{planck14} data release, with parameters: $\Omega_m = 0.307$, $\Omega_b = 0.048$, $\Omega_\Lambda = 0.693$ and $H_0 = 100 h$ km s$^{-1}$ Mpc$^{-1}$, where $h = 0.6777$.

A multi-mass particle ``zoom-in" technique \citep{jenkins13} was used to re-simulate the candidate haloes to higher resolution. The re-simulations were performed using the magneto-hydrodynamical code \textsc{arepo} \citep{springel10}. In this work we use the Level 4 resolution suite, where the typical mass of dark matter and baryonic particles are $3 \times 10^5 M_\odot$ and $5 \times 10^4 M_\odot$, respectively. Details regarding the subgrid galaxy formation processes are given in \cite{grand17}: these include critical processes such as star formation, stellar evolution and supernova feedback, a photoionizing UV background, metal line cooling, and the growth of supermassive black holes. The Auriga suite has been successful in reproducing a number of observational properties of both central discs and stellar haloes, including the rotation curves, stellar masses and star formation rates of discs \citep[e.g.][]{grand17, marinacci17}, and the kinematics and number density profiles of stellar haloes \citep[e.g.][]{deason17, monachesi18}. In this work we do not include Haloes 11 and 20 in our analysis, as they are both undergoing a merger at the present time.

The Milky Way analogues are defined as the central galaxies in the Auriga haloes, and the coordinate frame is based on the \textsc{subfind} algorithm \citep{davis85}. In this work, we only consider ``accreted" star particles (cf. \citealt{fattahi19}). These stars were bound to galaxies other than the main progenitors of the Milky Way analogues at the snapshot following their formation time. Thus, these stars mainly comprise the stellar debris from destroyed satellite galaxies. We choose to only include accreted stars for two main reasons: (1) there is little compelling evidence that the Milky Way stellar halo has significant contributions from stars born ``in-situ" \citep[e.g.][]{deason17, belokurov18, dimatteo18, haywood18} and (2) the presence of in-situ halo stars in simulations is strongly dependent on the subgrid galaxy formation physics and numerical resolution \citep[e.g.][]{zolotov09, cooper15}. Moreover, as recently found by \cite{monachesi18}, the inclusion of in-situ stars in the Auriga galaxies suites leads to stellar haloes that are substantially more massive and metal-rich than observations.

When examining the halo star kinematics around the solar radius $r_0 =8.3$ kpc, we rescale the phase-space distribution by the observed local circular velocity in the Milky Way, where $V_c(r_0) = 230 \kms$ \citep{eilers18}. The positions and velocities are multiplied by the scaling factor, $f=230/V_c(r_0)$, which ranges from $f \sim 0.75-1.4$.

\subsection{The definition of ``Escape Speed"}
\label{sec:def}
\begin{figure}
        \centering
        \includegraphics[width=\linewidth,angle=0]{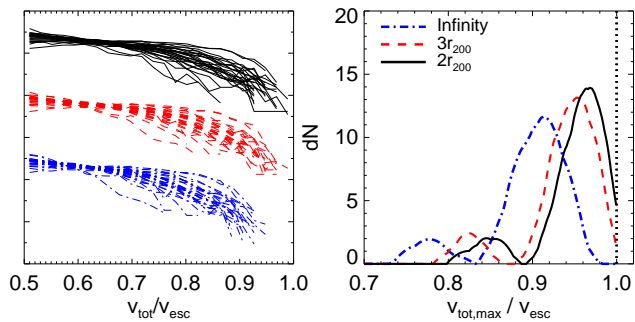}
        \caption{\textit{Left panel:} The velocity distribution of the Auriga stellar haloes relative to the escape speed. Different escape velocity definitions, shown by different colours, are shifted along the y-axis for clarity. \textit{Right panel:} The maximum speed reached by the stars relative to the escape velocity. Here, we only consider accreted stars in the radial range $4 < r/\mathrm{kpc} < 12$. The escape velocity is defined as escape to infinity (dot-dashed blue line), $3r_{200}$ (dashed red line) and $2r_{200}$ (solid black line), respectively. Note the curves are smoothed by an Epanechnikov kernel.}
          \label{fig:vtot_vesc}
\end{figure}

\begin{figure}
        \centering
        \includegraphics[width=\linewidth,angle=0]{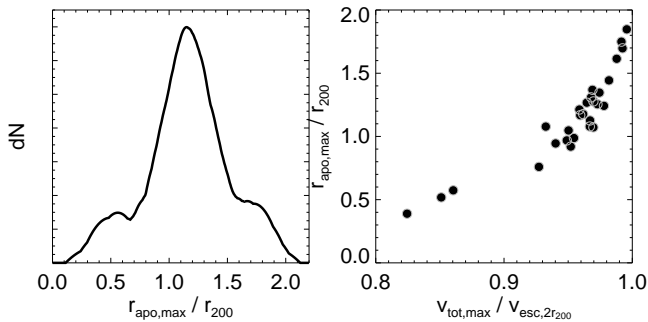}
        \caption{\textit{Left panel:} The distribution of maximum apocentres for the Auriga stellar haloes (for accreted stars between $4-12$ kpc) smoothed by an Epanechnikov kernel. The maximum radii are scaled by the virial radius, $r_{200}$.  \textit{Right panel:} The maximum apocentre as a function of maximum total velocity scaled by the escape velocity (defined as escape at $2r_{200}$.) Stars approaching the escape velocity typically have apocentres out to $\sim 1.5-2 r_{200}$.}
          \label{fig:rapo}
\end{figure}

\begin{figure}
        \centering
        \includegraphics[width=\linewidth,angle=0]{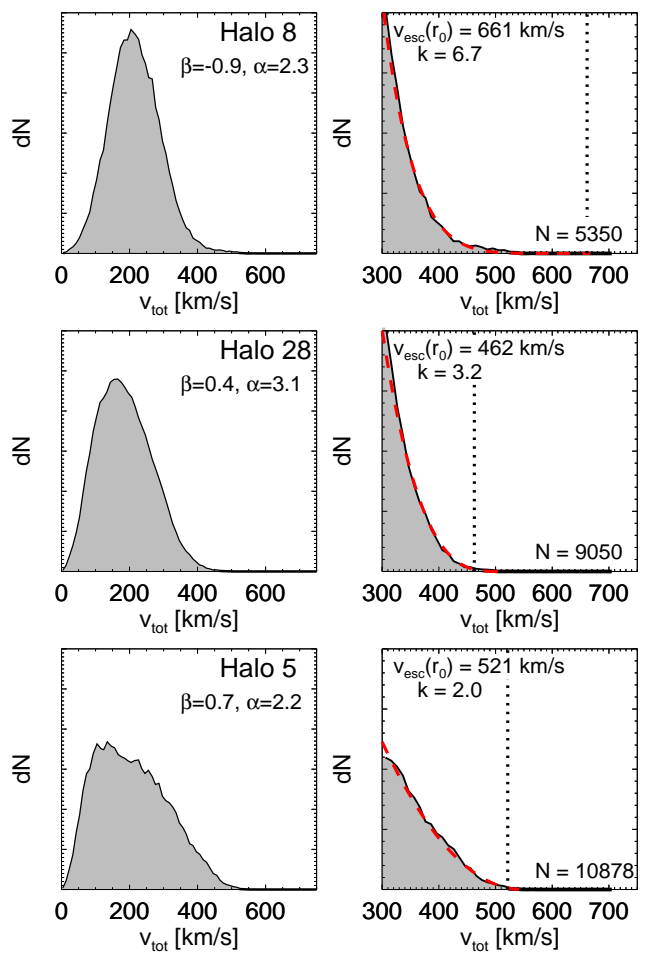}
        \caption{Total velocity distributions for three example haloes in the Auriga simulation suite. The left panels show the full distribution, and the right panels focus on the high velocity tail (with $v_{\rm tot} > 300 \kms$). The red dashed lines show a power-law fit to the high velocity tail, and the dotted line indicates the escape velocity. Note that we only consider accreted halo stars in the radial range $  4 < r/\mathrm{kpc} < 12$. The numbers in the bottom right corner indicate the number of star particles with $v_{\rm tot} > 300 \kms$.}
          \label{fig:vtot}
\end{figure}

The escape speed is defined as the velocity that a star requires to escape the gravitational field of a host halo. The simulated haloes are not isolated systems, so a limiting distance needs to be defined so that stars orbiting beyond this system can escape. In principle, this limiting distance is fairly arbitrary. However, the chosen distance should not underestimate the escape speed (i.e. to prevent stars being unrealistically unbound), but also should not reach far enough to permeate into the vicinity of neighbouring haloes. In the case of the Milky Way, a sensible choice is approximately half the distance to M31 (where $D_{\rm M31} \sim 800$ kpc). In addition, one would also like a definition of escape velocity which is plausible for stars in the solar neighbourhood. For example, if the limiting distance is too large then the maximum speeds' reached by the stars will not come close to the escape velocity. This consideration is important, as an intangible definition of the limiting distance will lead to an underestimate of the escape velocity, and hence the total mass.

\cite{piffl14} adopt an outer boundary of 3$r_{340}$, where the virial radius is defined relative to a density threshold of $340$ times the critical density. This leads to distances between 430 and 530 kpc. Note the definition of the virial radius used by \cite{piffl14} is not commonly used, but can easily be converted to the more standard definition of $200$ times the critical density ($r_{200}$, as used in this work): $r_{340} \approx 0.8r_{200}$, so $3r_{340} \approx 2.4r_{200}$. In comparison, \cite{smith07} use a slightly larger limiting distance of $3r_{200}$. 

In the left-hand panel of Fig. \ref{fig:vtot_vesc} we show the total velocity distributions of the Auriga stellar haloes relative to the escape velocity. In the right-hand panel we show the distribution of maximum speeds for each halo. Here, we consider accreted stars in the radial range $4 < r/\mathrm{kpc} < 12$. We use three definitions of escape velocity: relative to $2r_{200}$, $3r_{200}$, and the more unrealistic escape to infinity. We find that a limiting radius of $2r_{200}$ leads to stellar velocities approaching the escape velocity, but not passing it. Indeed, although not shown here, we find that closer limiting definitions of $r_{200}$ and $1.5r_{200}$ can lead to stars having velocities exceeding the escape velocity. In contrast, if we assume escape to infinity, the total velocities typically reach 90\% of the escape velocity. Although this may appear like a small decrement, for escape velocities of $\sim 500 \kms$ this can lead to underestimates of $\sim 50 \kms$. In the remainder of this work we choose $2r_{200}$ as the limiting radius in our fiducial definition of $v_{\rm esc}$. This radius ranges from $2r_{200} \sim~400-500$ kpc in the Auriga haloes. Conveniently, this definition also approximately coincides with the halfway distance to M31.

We explore our definition of the limiting radius further by examining the apocentres of the high velocity stars. To approximately estimate the apocentres, we calculate the Energy ($E_0$) and total angular momentum ($L_0$) of stars with $v_{\rm tot} > 300 \kms$, and find the radii where $\Phi(r) +L^2_0 / 2 r^2 = E_0$ \citep[see][chapter 3]{bt87}. Here, we only consider stars in the radial range $4 <  r/\mathrm{kpc} <12$ at $z=0$. To estimate the potential of the simulated haloes, we assume spherical symmetry and consider all particles in the radial range $0 < r/\mathrm{kpc} < 600$. For each halo, we find the maximum apocentre, which generally coincides with the more extreme $v_{\rm tot}$ values. In the left-hand panel of Fig. \ref{fig:rapo} we show the distribution of maximum apocentres scaled to the virial radius, $r_{200}$. There is a wide range of radii, but typically these lie at $\sim 1-1.5r_{200}$.  In the right-hand panel of Fig. \ref{fig:rapo} we show how these maximum apocentres relate to the maximum velocities. Typically, stars with velocities approaching the escape speed have apocentres of $\sim 1.5-2 r_{200}$. Thus, this exercise shows that our choice of $2r_{200}$ as an outer boundary is also appropriate based on the orbits of the high velocity stars.

Figures \ref{fig:vtot_vesc} and \ref{fig:rapo} show that there is a great deal of variation between the Auriga haloes. Indeed, some stellar velocity distributions reach right up to the escape velocity, whilst others are truncated well below it. This is related to the varying forms of the high velocity tails, which, as we showed in the previous Section, are dependent on the properties of the halo stars, such as their velocity anisotropy and radial density profile. Indeed, a significant advantage of using the Auriga suite is that the number of haloes ($N=28$ used in this work) is sufficient to probe a wide range of assembly histories (cf. \citealt{smith07} and \citealt{piffl14} who used four and eight haloes in their analyses, respectively). This is particularly important if the Milky Way's accretion history is atypical. However, before proceeding we caution that although the Auriga suite has significantly more high resolution Milky Way-like haloes than previous simulations, this does not guarantee that the assembly histories of the simulated haloes are sufficiently close to that of the Milky Way. Indeed, our findings are, like others, limited by the variety of assembly histories present in Auriga. Nonetheless, we believe that the range of accretion histories of the Auriga haloes presents a fair sampling of the halo-to-halo scatter at this mass range, and, at present, is the best equipped simulation suite for this work.

\subsection{High velocity tails in Auriga}

In this Section, we explore the high velocity tails of the accreted stellar haloes in the Auriga simulations. Throughout, we consider stars in the radial range $4 < r/\mathrm{kpc} < 12$, which brackets the solar radius of the Milky Way. In Fig. \ref{fig:vtot} we show three example velocity distributions. The high velocity tails are highlighted in the right-hand panels, and the red-dashed line shows a fit of the form Eqn. \ref{eq:lt90} to stars with $v_{\rm tot} > 300 \kms$. Here, we have fixed the escape velocity --- defined with a limiting radius of $2r_{200}$ --- and allowed $k$ to be a free parameter. Note that the escape velocity varies as a function of radius, so each star at a given radius has a slightly different escape velocity. In Fig. \ref{fig:vtot} we indicate the best-fit $k$ value, and the escape velocity at $r = r_0=8.3$ kpc. These examples bracket cases with steep velocity tails (e.g. Halo 8, $k=6.7$) and shallow velocity tails (e.g. Halo 5, $k=2.0$).

\begin{figure}
        \centering
        \includegraphics[width=\linewidth,angle=0]{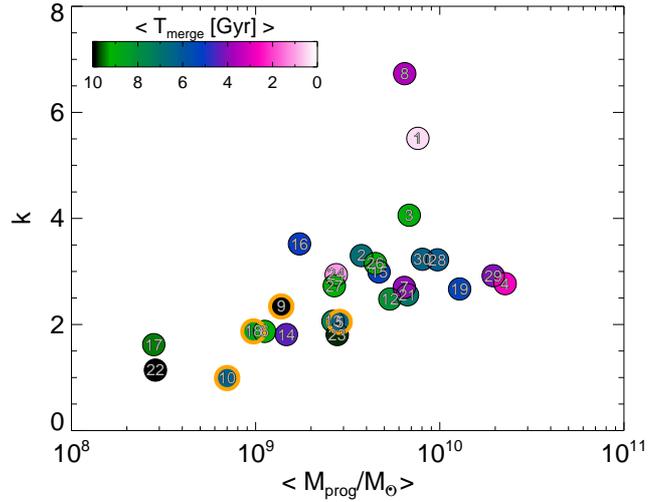}
        \caption[]{The power-law slope of the high velocity tail of accreted halo stars against the most massive progenitor contributing to the velocity distribution.  Here, we consider stars in the radial range $4 < r/\mathrm{kpc} < 12$. The points are coloured according to the merger time of the dwarf galaxy, and the halo ID number is indicated in grey. Note that the haloes with prominent ``sausage'' components (highlighted in orange --- see \citealt{fattahi19} Fig. 3) have low $k$ values.}
          \label{fig:k_mprog}
\end{figure}

In Fig. \ref{fig:k_mprog} we show the derived $k$ values for each Auriga halo as a function of the median dwarf progenitor mass of the accreted stars in the radial range $4 < r/\mathrm{kpc} < 12$. Note that, in most cases, there are one or two progenitors that contribute the majority of halo stars (see e.g. \citealt{fattahi19}). The circle points are coloured according to the median lookback time that the stars became bound to the Milky Way's main progenitor (rather than the dwarf progenitor). This figure shows that recent, massive accretion events lead to larger $k$ values than earlier, less massive events. We also indicate, with the orange circles, the four haloes with very prominent ``sausage" components --- i.e. with highly anisotropic velocity distributions --- found by \cite{fattahi19}. These have low values of $k$, with $k \lesssim 2.5$ (see below). Fig. \ref{fig:k_mprog} shows that the variation of $k$ depends on the assembly history of the haloes. Thus, as alluded to in the previous Section, our knowledge of the formation of the inner Milky Way stellar halo provides a key constraint on $k$. Recent results from \gaia\ suggest that the inner halo was built from the disruption of an SMC or LMC mass ($M_{\rm star} \sim 10^9M_\odot$) dwarf galaxy at early times ($T \sim 8-10$ Gyr) \citep[e.g.][]{belokurov18, helmi18}, and thus, based on Fig. \ref{fig:k_mprog}, low values of $k < 2.5$ are preferred.
 
\begin{figure*}
        \centering
        \includegraphics[width=17cm,angle=0]{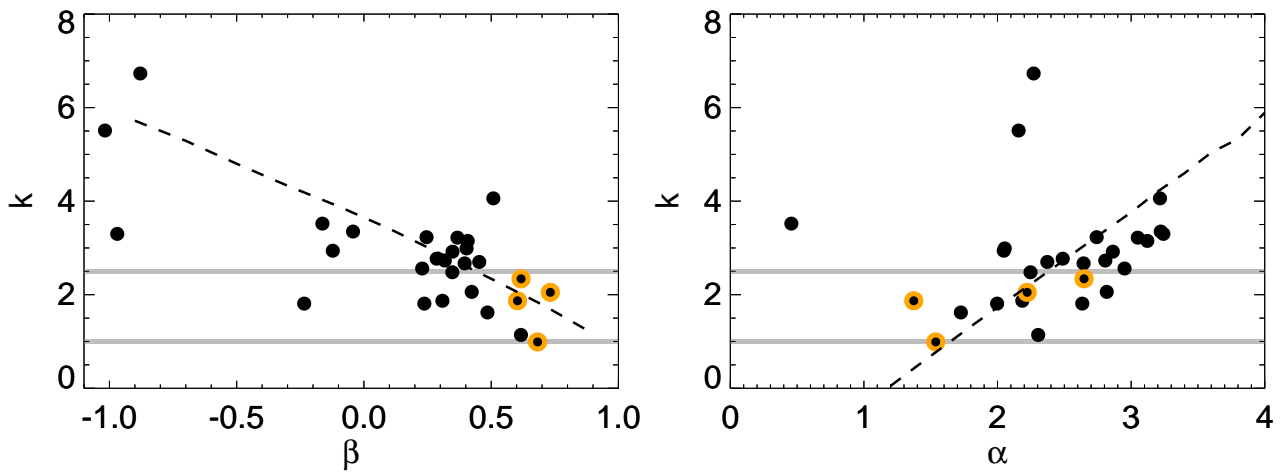}
        \caption{ The power-law slope of the high velocity tail in the Auriga haloes as a function of velocity anisotropy ($\beta$, left panels) and stellar halo density slope ($\alpha$, right panels). The most prominent ``sausage" haloes in the Auriga suite are highlighted in orange. Note all parameters are calculated within the radial range $4 < r/\mathrm{kpc} < 12$. The black dashed lines indicate the relation between $k$ and $\beta \, (\alpha)$ predicted by the power-law dfs. Here, we have fixed $\alpha \, (\beta)$ and $\gamma$ to the median values of the simulated haloes. As predicted by the analytical dfs, the tails of the velocity distributions are shallower when the velocity anisotropy is strongly radial and/or the stellar halo density is relatively shallow. The thick grey lines indicate the range of $k$ appropriate for stellar haloes with strongly radial velocity anisotropy.}
          \label{fig:k_beta_alpha}
\end{figure*}

We can explore in more detail how $k$ depends on the stellar halo properties by analysing the phase-space distribution of the stars. In Fig. \ref{fig:k_beta_alpha} we show how $k$ depends on the velocity anisotropy ($\beta$, left panel) and the power-law slope of the stellar halo density ($\alpha$, right panel). Note that both of these quantities ($\beta$ and $\alpha$) are measured within the radial range $4 < r/\mathrm{kpc} < 12$. As we found in the idealised power-law distribution function models (see Sec. \ref{sec:dfs}), higher $\beta$ and/or lower $\alpha$ values lead to lower values of $k$. The dashed black lines indicate the predicted relations from the analytical dfs, where $\gamma$ and $\alpha$ or $\beta$ is fixed to the median values of the simulated haloes ($\gamma=0.3, \alpha=2.5, \beta=0.35$). Remarkably, these predictions agree well with the simulations!

The four haloes with prominent ``sausage" components are again highlighted in orange. We also indicate with the thick grey lines the range of $k \in [1.0,2.5]$ appropriate for stellar haloes with strongly radial velocity anisotropy. Figures \ref{fig:k_mprog} and \ref{fig:k_beta_alpha} illustrate that, although there is a relatively wide range of $k$ values in the simulations ($ 1 \lesssim k \lesssim 7$), the form of the high velocity tail is correlated with the stellar halo properties. Thus, rather than bracket the range predicted by the simulations, which covers a wide range of assembly histories, we can provide a more stringent constraint on $k$ from our observational data. Thus, in the following Section, when we measure the local Galactic escape speed, we impose $1.0 < k < 2.5$. This range of $k$ encompasses the values we found in the Auriga simulations when $\beta \sim 0.7$, and also brackets the predicted $k$ value from the analytical power-law dfs when $\beta=0.7, \alpha=2.5$.

\subsubsection{Constraining the local escape velocity}
\begin{figure}
        \centering
        \includegraphics[width=\linewidth,angle=0]{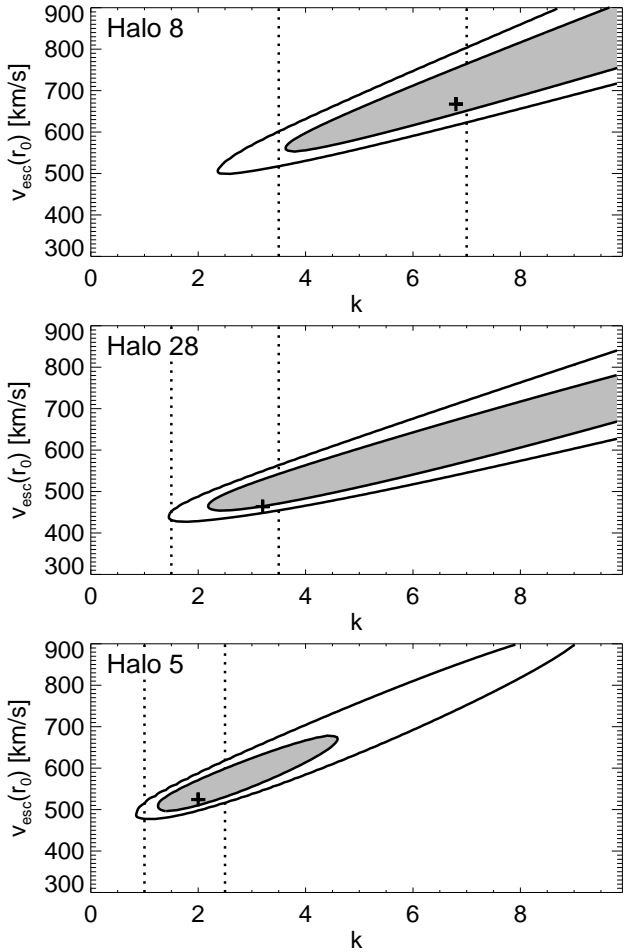}
        \caption{The 2D confidence contours in the $k$ and $v_{\rm esc}(r_0)$ space for three example Auriga haloes (see also Fig. \ref{fig:vtot}). We have marginalised over the radial power-law slope of the escape velocity ($\gamma$), and the contours show the $1-$ (grey filled) and $2-\sigma$ (solid line) confidence regions. Here, we have randomly chosen $N=240$ star particles in the radial range $4< r/\mathrm{kpc} < 12$ with $v_{\rm tot} > 300 \kms$, and include a random error on the total velocities of $30 \kms$. This approximately mimics the sample size and uncertainties in the \gaia\ data (see Section \ref{sec:gaia}).   The degeneracy between $k$ and $v_{\rm esc}(r_0)$ is clear. Moreover, with smaller samples sizes and/or relatively large velocity errors, the degeneracy becomes even more pronounced. The dotted lines indicate the approximate range of $k$ predicted based on the velocity anisotropy of the halo stars --- the addition of this constraint can narrow down the allowed region of $v_{\rm esc}(r_0)$ substantially. }
\label{fig:vesc_k_sims}
\end{figure}

We end this Section by illustrating the importance of $k$ in determining an accurate Galactic escape speed. Here, we perform the maximum likelihood analysis described in Section \ref{sec:lt90} to the simulation data. Here, $k$, $\gamma$ and $v_{\rm esc}(r_0)$ are free parameters. To mimic the approximate status of the observational data, we randomly choose $N=240$ star particles in the radial range $4< r/\mathrm{kpc} < 12$ with $v_{\rm tot} > 300 \kms$, and include a Gaussian error on the total velocities with $\sigma = 30 \kms$. Note this exercise is for illustration rather than quantification of the observational results (see Section \ref{sec:gaia}). In Fig. \ref{fig:vesc_k_sims} we show the 2D confidence contours in the $k$ and $v_{\rm esc}(r_0)$ space for the three example Auriga haloes shown in  Fig. \ref{fig:vtot}. Here, we have marginalised over the power-law slope of the potential ($\gamma$), but note that this parameter is generally poorly constrained when there is a limited radial range and small number of tracers (see Fig. \ref{fig:gaia_vesc_slope}). Fig. \ref{fig:vesc_k_sims} shows that, although the true $k$ and $v_{\rm esc}(r_0)$ values are contained within the $1-\sigma$ confidence regions (plus symbols), there is a strong degeneracy between $k$ and $v_{\rm esc}(r_0)$ , such that the escape velocity varies by hundreds of $\kms$ when $k$ is unknown. The dotted lines indicate the approximate range of $k$ predicted based on the velocity anisotropy of the halo stars (see Fig. \ref{fig:k_beta_alpha}) --- this prior knowledge can substantially narrow down the  allowed range of $v_{\rm esc}(r_0)$ values. Note that we impose a range of $k$, rather than a fixed value, to account for the scatter in $k$ at fixed $\beta$.

For several reasons, the case of our own Milky Way appears rather fortuitous! First, the currently accepted origin of the inner stellar halo --- namely from the debris of one massive dwarf, accreted several Gyr ago --- suggests that the majority of the stellar halo material, at least near the solar vicinity, is well phase-mixed. Second, as mentioned previously, our knowledge of the halo stars' orbits  in the solar vicinity places a constraint on $k$, with $1.0 < k < 2.5$. Third,  the fact that the Milky Way likely has a low $k$ value means that the high velocity stars can more strongly constrain the escape velocity. For example, if $k=1$, the high velocity tail linearly declines to a truncation at $v_{\rm esc}$. Thus, in this case, the fastest star in the sample is likely very close to the escape velocity. In contrast, if $k$ is high, a long, poorly populated tail extends to the escape velocity, and thus the escape velocity is more difficult to constrain.

On that optimistic note, we end this Section exploring the Auriga simulations, and proceed to constrain the local Galactic escape speed using \gaia\ data.

\section{The Galactic escape speed from \emph{Gaia} DR2}
\label{sec:gaia}
In this Section, we apply the LT90 formalism described in Section \ref{sec:lt90} to \gaia\ data release 2 (DR2, \citealt{gaia18}). We use the information gleaned from the simulations to help constrain the escape velocity by applying a prior on the $k$ value, which is tailored for our own Milky Way galaxy.

\subsection{\gaia\ DR2 data}
We select stars from \gaia\ DR2 with parallax, proper motion and radial velocity information. We apply the same quality flags as \cite{marchetti18} and \cite{monari18} to make sure our sample is free from spurious objects. In addition,  we only include stars with re-normalised unit weight error, $\mathrm{RUWE} < 1.4$ \citep{ruwe}, which ensures stars with unreliable astrometry are excluded. Our estimate of $v_{\rm esc}(r_0)$ is sensitive to the fastest moving stars, hence we restrict our analysis to stars with accurate parallax measurements, with $0 <\sigma(\varpi)/\varpi < 0.1$. To estimate distances, we use the procedure outlined in \cite{mcmillan18}, which uses a prior designed to apply to the \gaia\ data with radial velocities. We adapt the method\footnote{The code from \cite{mcmillan18} is available here: \url{https://github.com/PaulMcMillan-Astro/GaiaRVStarDistances}} to only include a prior relevant for a halo population. In practice, this means only considering a halo density component (rather than multiple Galactic components), and assuming a flat age and metallicity prior. We assume a power law slope with index $-2.5$ for the halo stars, in agreement with the most recent constrains for the density profile of the inner halo  \citep[e.g.][]{faccioli14, pila15}. In our analysis, we only include stars in the immediate solar vicinity with $D < 3$ kpc: this cut ensures our distances are dominated by the parallax information rather than the prior. Finally, to avoid any contamination from disc stars, we only consider counter-rotating stars (cf. \citealt{monari18}). Our final sample of stars is $N \sim 2300$, of which $N \sim 240$ have $v_{\rm tot} > 300 \kms$. With future \gaia\ data releases we can be less restrictive, and explore a wider range of distances. Here, we focus on a local sample in order to robustly determine $v_{\rm esc}(r_0)$.

The distances, proper motions and radial velocities are converted to Galactocentric coordinates, assuming a circular velocity of $v_c(r_0)= 230 \kms$ \citep{eilers18} at the position of the Sun ($r_0=8.3$ kpc), and a peculiar solar motion of ($U_\odot, V_\odot, W_\odot) = (11.1, 12.24, 7.25) \kms$ \citep{schonrich10}. If the adopted circular velocity is lower or higher by 10 $\kms$ then our derived total velocities are only mildly affected, and our measured escape velocity is not significantly changed. We propagate errors in our analysis using a Monte-Carlo technique. Samples are generated $N= 1000$ times with proper motions, distances and radial velocities drawn from their respective error distributions. The data is resampled with replacement (cf. \citealt{smith07}), and in each iteration we only consider stars with $v_{\rm tot} > 300 \kms$, $v_\phi < 0 \kms$ and $D < 3$ kpc. We employ a brute force grid-based method to estimate the likelihood values, with uniform grids in the range $k \in [0,10]$, $v_{\rm esc}(r_0) \in [400,900]$ and $\gamma \in [0,1]$. 

\subsection{Results}

\begin{figure}
        \centering
        \includegraphics[width=\linewidth,angle=0]{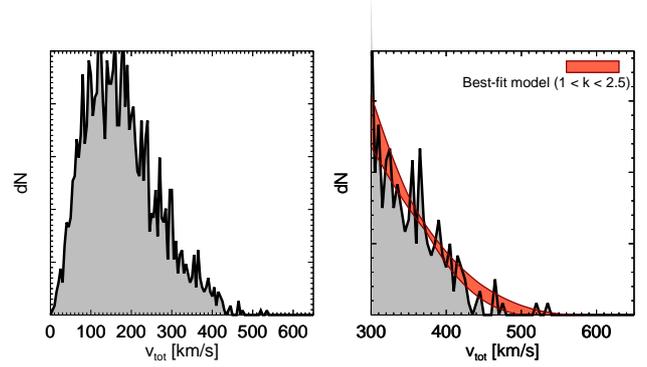}
        \caption[]{The velocity distribution of $N \sim 2300$ counter-rotating stars in the \gaia\ data release 2 catalogue. These stars have measured proper motions, radial velocities and parallaxes. We select stars within 3 kpc of the solar neighborhood, with less than 10\% parallax errors. In the right-hand panel, the red line-filled polygon shows the best-fit model to the high velocity tail.}
          \label{fig:gaia_vtot}
\end{figure}

\begin{figure*}
        \centering
        \includegraphics[width=\linewidth,angle=0]{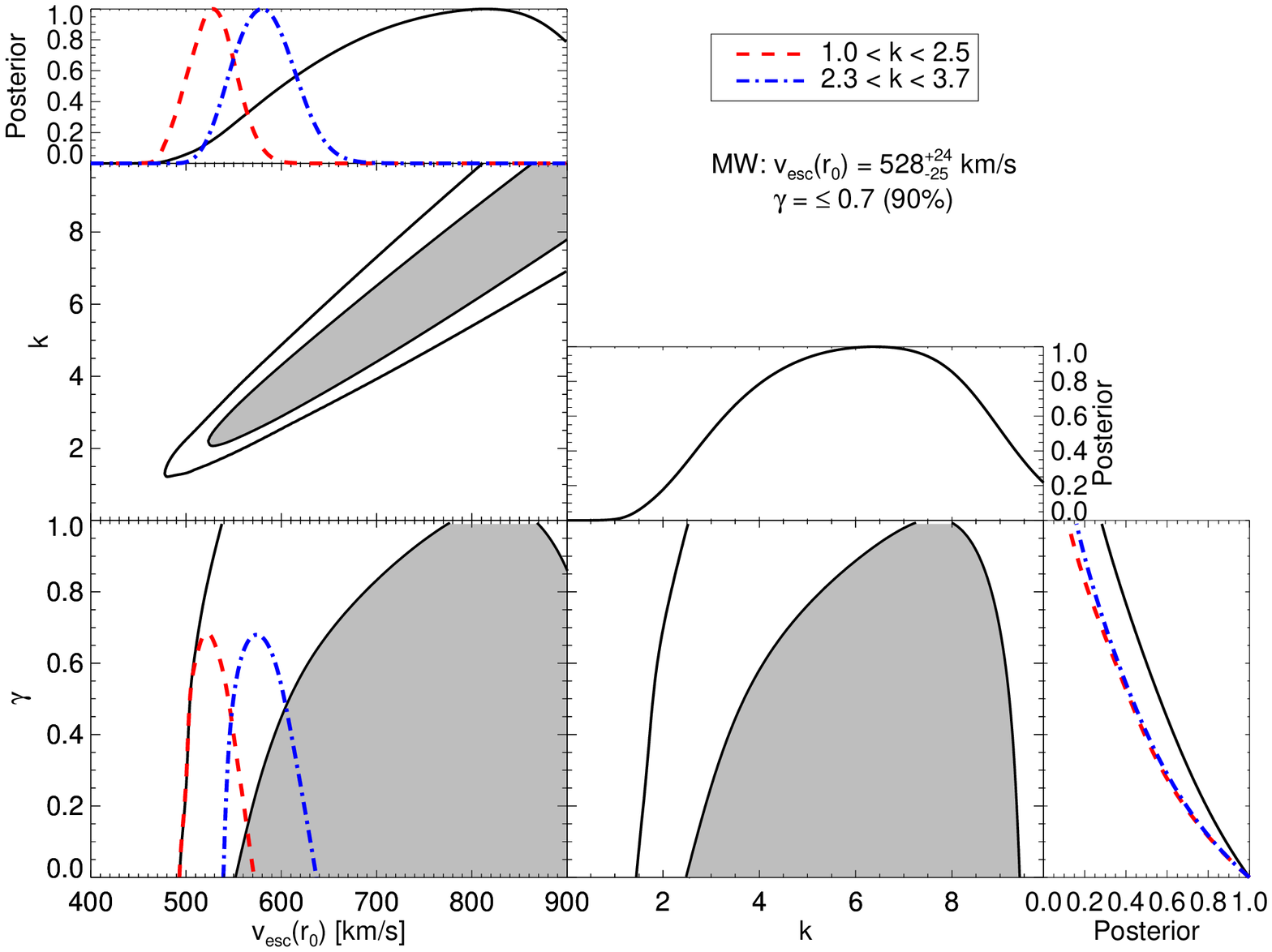}
        \caption[]{The results of applying our likelihood analysis to the \gaia\ data with $v_{\rm tot} > 300 \kms$. Here, the degeneracy between $v_{\rm esc}$ and $k$ is clear. When we adopt a prior of $1 < k < 2.5$ (red dashed line), appropriate for the strongly radial orbits observed in the solar neighbourhood, we find $v_{\rm esc}(r_0) = 528^{+24}_{-25} \kms$. Note that adopting the same prior as \cite{monari18} and \cite{piffl14}, $2.3 < k < 3.7$ (blue dot-dashed line), results in a larger escape velocity: $v_{\rm esc}(r_0) = 580^{+32}_{-32}$. We find little evidence for strong radial variation in $v_{\rm esc}$ over the range we're probing (i.e. $\gamma \sim 0$), with $\gamma \leq 0.7$ with 90\% confidence. }
          \label{fig:gaia_vesc_slope}
\end{figure*}

The total velocity distribution of the \gaia\ data is shown in Fig. \ref{fig:gaia_vtot}, where stars with $v_{\rm tot} > 300 \kms$ are shown in the right-hand panel. When we adopt a flat prior of $ 1 < k < 2.5$, which is appropriate for the highly eccentric orbits in the solar vicinity, the best-fit model is indicated by the red band. The width of the band indicates the $90\%$ confidence region.

The confidence regions for $k$, $v_{\rm esc}(r_0)$ and $\gamma$ are shown in Fig. \ref{fig:gaia_vesc_slope}. The filled grey region and solid black line shows the $1-$ and $2-\sigma$ confidence intervals, respectively. Here, we have assumed flat priors for $k$ and $\gamma$ and employed a Jeffrey's prior for $v_{\rm esc}(r_0)$. We show the posterior distributions for each parameter in the inset panels. The degeneracy between $k$ and $v_{\rm esc}(r_0)$ is clear, as seen in the previous Section (and earlier work by \citealt{smith07} and \citealt{piffl14}). The red and blue lines illustrate the effect of a prior on $k$. Specifically, the dashed red line applies our new prior --- based on the orbits in the solar neighbourhood, and calibrated on the Auriga simulations --- of $1 < k < 2.5$. For comparison, we also show the prior adopted by \cite{piffl14} and \cite{monari18}, which is also based on cosmological simulations: $2.3 < k < 3.7$. In these works, the prior spans the range of $k$ values found in simulations. However, our adopted prior is tailored towards the highly eccentric stars in the Milky Way, which leads to lower $k$ values.

Assuming $ 1 < k < 2.5$ we find $v_{\rm esc}(r_0) = 528^{+24}_{-25} \kms$. This value is lower than the recent determination by \cite{monari18} using \gaia\ DR2 data. However, the reason for this difference is owing to the prior information on $k$. If we adopt the \cite{piffl14} prior, we find $v_{\rm esc} = 580^{+31}_{-31} \kms$, which is in excellent agreement with \cite{monari18}. Note that our error bars are smaller than \cite{monari18} because we do not use narrow distance bins, but rather use all the data and allow for a radially varying escape velocity. Our estimate of the local escape velocity is in good agreement with the values found by \cite{smith07}, \cite{piffl14} and \cite{williams17}, who used line-of-sight velocity data from RAVE and SDSS to derive $v_{\rm esc}$. However,  it is curious that these works find a similar escape velocity, as in all cases larger values of $k$ were adopted --- which should, presumably, bias towards larger $v_{\rm esc}$ values. These works used samples of high latitude stars with line-of-sight velocity measurements only, and thus if there was any flattening in the stellar halo distribution in the $z$ direction, the total speed estimates based on the line-of-sight velocities could be biased low. In particular, we now know that the inner stellar halo is significantly flattened \citep[e.g.][]{iorio18}, and the highly eccentric orbits that dominate the high velocity tail are generally confined close to the Galactic plane \citep[e.g.][]{myeong18}. Thus, we suggest that the line-of-sight analysis performed by \cite{smith07}, \cite{piffl14} and \cite{williams17} would underestimate $v_{\rm esc}$ if they used the correct $k$ prior. Instead, we postulate that the underestimate due to the flattened halo combined with a bias towards larger $k$ values has conspired to give an answer consistent with our results!

Finally, we remark that our constraint on $\gamma$ is weak, with $\gamma=0$ consistent with the data. This is unsurprising given that we do not explore an extensive distance range. However, when we can probe to larger distances with future \gaia\ data releases. our methodology can be used to also constrain $\gamma$, and hence the slope of the potential.

\subsubsection{Bound or unbound?}
The local escape velocity has often been used to ascertain whether or not stars with extreme velocities are bound to the Milky Way. Indeed, there exists a population of stars with velocities exceeding the escape velocity, which are often labeled as ``hyper-velocity stars"  or ``hyper-runaway" stars \citep[see e.g.][]{brown15}. There are several plausible mechanisms that may have formed these fast moving stars, including interactions with the central supermassive black hole \citep[e.g.][]{hills88}, ejection from the Large Magellanic Cloud \citep[e.g.][]{boubert16}, dynamical encounters between star clusters \citep[e.g.][]{leonard90},  and supernova explosions in stellar binary systems \citep[e.g.][]{portegies00}. However, while there exist a small number of extreme cases, the origin of many stars with high velocities are uncertain, as their velocities straddle the boundary of the Galactic escape velocity. Thus, an accurate measure of the escape velocity is vital in order to determine the origin of the fastest moving stars.

Recently, several works have used \gaia\ DR2 data to compile samples of candidate stars with extreme velocities \citep[e.g.][]{bromley18, hattori18, marchetti18}. However, based on both orbital and chemical arguments, \cite{boubert18} and \cite{hawkins18} argue that the vast majority of these candidates are likely bound to the Milky Way, and comprise the high velocity tail of the stellar halo. Our constraint on the local escape velocity, coupled with the observational errors, agrees with this hypothesis. More accurate constraints on the escape velocity, and hence the Galactic potential, will allow a more stringent classification of the origin of the apparently extreme stars. Moreover, while \gaia\ DR2 is a giant leap forward in Galactic astronomy, future data releases will limit the number of statistical outliers, which are inevitable with these early \gaia\ data releases.

The evidence that several of the fastest moving stars occupy the high velocity tail of the stellar halo reinforces the finding of this work. Namely, that the high velocity tail of the local stellar halo is well populated owing to the significant radial velocity anisotropy of the halo stars. Indeed, if the velocity distribution was more sharply truncated, as we saw in some of the Auriga haloes, then we would see a less significant population of (bound) high velocity stars.

\begin{figure*}
        \centering
        \includegraphics[width=16cm,angle=0]{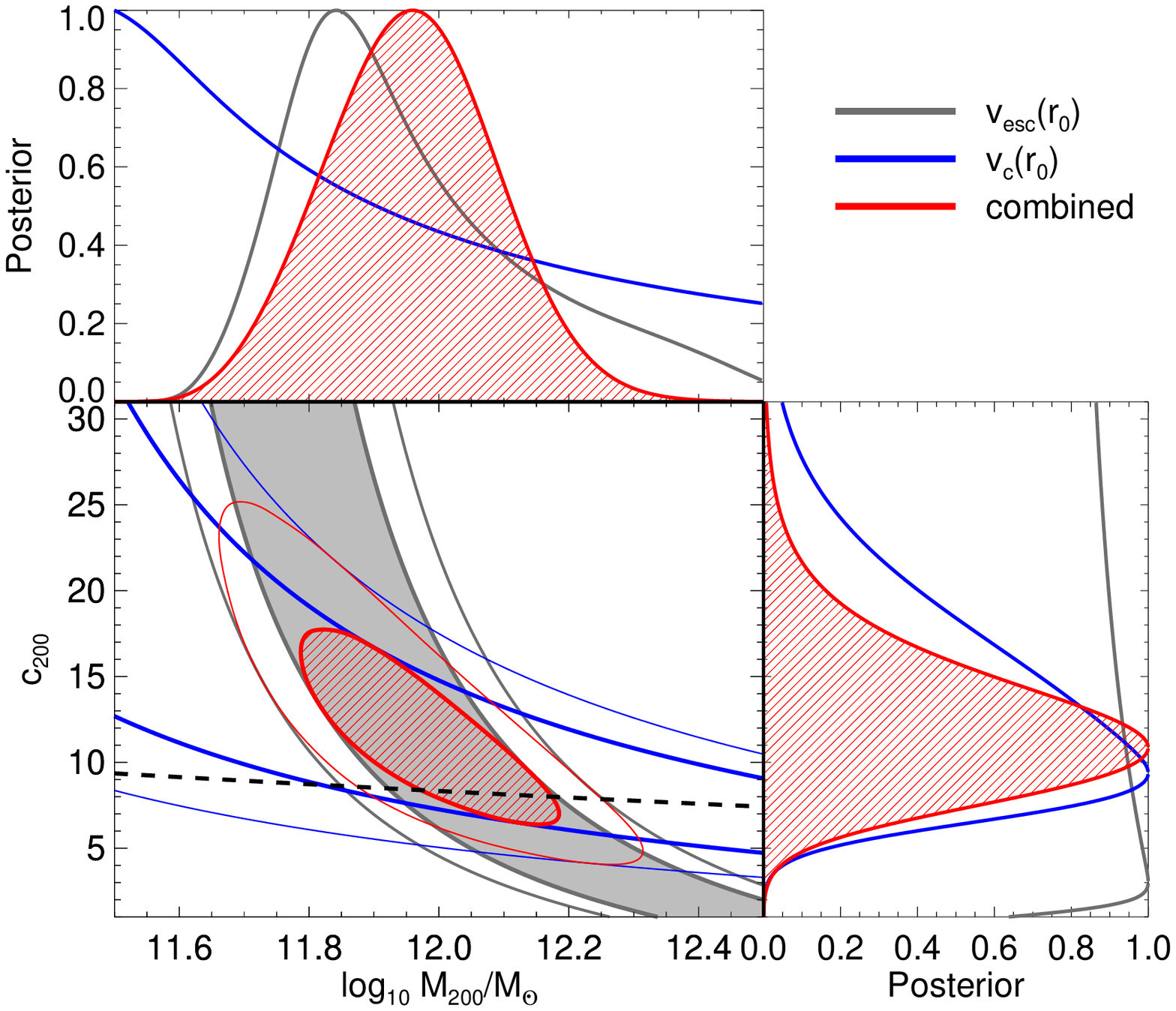}
        \caption[]{The derived NFW halo parameters from our escape velocity measurement. Here, we assume a bulge and 2-component disc potential as given in \cite{pouliasis17} (also used in \citealt{eilers18}). The gray filled contour shows the 68\% confidence, and the solid gray line shows the 95\% confidence region. The blue contours uses constraints on the local circular velocity: $v_c(r_\odot) = 230 \pm 10 \kms$. The red contours indicate the combined constraint. The black dashed line indicates the mass-concentration relation from \cite{dutton14}. In the top panel and right-hand panel we show the $1D$ posterior distributions for $M_{200}$ and $c_{200}$, respectively. Our derived dark halo mass is: $M_{200} = 0.79^{+0.45}_{-0.17} \times 10^{12}M_\odot$ (escape velocity only), $M_{200} = 0.91^{+0.31}_{-0.24} \times 10^{12}M_\odot$ (escape velocity and circular velocity).}
          \label{fig:nfw}
\end{figure*}

\section{Total Milky Way mass}
\label{sec:mass}

The local escape velocity is a direct measure of the gravitational potential. Historically $v_{\rm esc}$ has been regarded as the velocity required to escape to infinity, so $v_{\rm esc}(r) = \sqrt{2\Phi(r)}$, however, in practice, this definition is unrealistic. Instead, one needs to define a limiting radius beyond which a star is considered unbound (or cannot fall back onto the galaxy). In Section \ref{sec:sims}, we found that the appropriate limiting radius in the Auriga haloes is $\sim 2 r_{200}$, thus when we convert our estimated escape velocity to a total mass estimate we need to consider $v_{\rm esc}(r_0)=\sqrt{2\left(\Phi(r_0)-\Phi(2r_{200}\right))}$.

From this definition, we can constrain the dark matter halo parameters from our estimated escape velocity. We assume an NFW \citep{nfw96,nfw97} profile and let $M_{200}$ and $c_{200}$ be free parameters. We fix the baryonic components of the Galactic potential, adopting Miyamoto-Nagai profiles \citep{miyamoto75} for the thin and thick discs, and a spherical Plummer potential \citep{plummer} for the bulge. We use the parameters of the enclosed mass, scale-lengths and scale-heights from Model I in \cite{pouliasis17}. We vary $M_{200}$ and $c_{200}$ uniformly in the ranges $\mathrm{log}(M_{200}) \in [11.5,12.5]$ and $c_{200} \in [1,30]$, respectively. To derive the NFW parameters, we use the posterior values for $v_{\rm esc}(r_0)$ derived in the previous Section, after marginalising over $\gamma$ and $k$, and assuming $1< k < 2.5$.

The grey contours in Fig. \ref{fig:nfw} show the confidence intervals for the NFW parameters (grey filled is $1-\sigma$, grey line is $2-\sigma$). We also show with the blue lines (thicker line is $1-\sigma$, thinner line is $2-\sigma$) the constraints on $M_{200}$ and $c_{200}$ assuming the circular velocity at the position of the Sun is $v_c(R_0) = 230 \pm 10 \kms$ \citep{eilers18}. The combined constraint from $v_{\rm esc}$ and $v_c$ is shown with the red contours. Interestingly, the $v_{\rm esc}$ and $v_c$ constraints are perpendicular to each other in the $M_{200}$, $c_{200}$ plane: this is because the escape velocity contains information about the potential exterior to the solar radius, whereas the circular velocity mainly depends on the mass interior.  This results in a stronger constraint on $M_{200}$ and $c_{200}$ when the $v_{\rm esc}$ and $v_c$ measurements are combined, and we find $M_{200} = 0.91^{+0.31}_{-0.24} \times 10^{12}M_\odot$ and $c_{200} = 10.9^{+4.4}_{-3.3}$. Note this relates to a \emph{total} mass measurement, including the baryonic mass, of $M_{200, \rm tot}= 1.00^{+0.31}_{-0.24} \times 10^{12}M_\odot$.

The black dashed line in Fig. \ref{fig:nfw} shows the mass-concentration relation derived by \cite{dutton14} for dark matter only simulations. For our estimated dark matter mass, $M_{200} = 0.9 \times 10^{12}M_\odot$, the \cite{dutton14} relation predicts a concentration of $c_{200}=8.4$. Our derived value of $c_{200} = 10.9^{+4.4}_{-3.3}$ is higher than the theoretical prediction, but agrees within the 1-$\sigma$ errors. Moreover, our derived concentration is in good agreement with recent constraints in the literature \citep[e.g.][]{callingham19}. If, however, we fix the concentration in our analysis to the \cite{dutton14} prediction we find a total mass measurement of $M_{200, \rm tot}= 1.29^{+0.22}_{-0.22} \times 10^{12}M_\odot$. Note that we get very similar results if we adopt the mass-concentration relations derived by \cite{schaller15} and \cite{ludlow16}.

Our prior on $k$ prior strongly influences the derived local escape velocity, and thus also the estimated halo mass. For example, if we adopt the same prior on $k$ as \cite{monari18} then our dark halo mass estimate is $M_{200} = 1.5 \times 10^{12}M_\odot$ (or $M_{200} = 1.7 \times 10^{12}$ if the \citealt{dutton14} mass-concentration relation is assumed). These values are in good agreement with \cite{monari18}, which is reassuring as they also use \gaia\ DR2 in their analysis. Interestingly, although our derived escape velocity is similar to \cite{piffl14}, they find a more massive Milky Way halo, with $M_{200, \rm tot} \sim 1.6 \times 10^{12}M_\odot$. However, we find that the main cause of this discrepancy is the mass-concentration relation assumed by \cite{piffl14}. They use the \cite{maccio08} mass-concentration as a prior, which is based on the WMAP5 cosmology. However, in the Planck cosmology (as used by \citealt{dutton14}) the concentrations are 20\% higher. Thus, by adopting the \cite{dutton14} mass-concentration relation based on Planck, our mass estimates are $\sim 20\%$ lower. In addition to the different mass-concentration relation, \cite{piffl14} also adopt a lower circular velocity, $v_c = 220 \kms$. This also leads to a slightly higher mass estimate (see Fig. 13 in \citealt{piffl14}), but, as we assume a 10 $\kms$ error in the local circular velocity, this difference is subsumed into the mass uncertainty.

Finally, we also comment on the limiting radius that defines the escape velocity. In this work, we find that $2r_{200}$ is the most appropriate choice (see Section \ref{sec:def}). However, if we adopted larger radii (i.e. $\sim 2.4-3 r_{200}$, cf. \citealt{smith07, piffl14}) our mass estimates would be slightly lower. For example, a limiting radius of $3r_{200}$ reduces our total mass estimate by $\sim 8$\%. This lower mass is due to the limiting radius being overestimated, and hence the estimated escape velocity is lower than the true velocity needed to escape. Thus, the choice of limiting radius is an important consideration when relating local escape velocity measurements to constraints on the total mass.

Since the first astrometric \gaia\ data release (DR2) several works have provided updated estimates of the total Milky Way mass \cite[e.g.][]{eadie18, malhan18, watkins18, callingham19, posti19, vasiliev19}. The majority of these use globular clusters or stellar streams confined within $\sim 50$ kpc, so a total mass estimate out to the virial radius requires an extrapolation. \cite{watkins18}, \cite{posti19} and \cite{vasiliev19} find $M_{\rm vir, tot} =1.2-1.5 \times 10^{12}M_\odot$ using the dynamics of globular clusters in the inner halo, and extrapolate to the virial radius using mass-concentration relations. Here, these authors have used the definition of virial radius adopted by \cite{bryan98} and \cite{klypin02}; the mass is defined within $340 \Omega_M$ ($\approx 100$) times the critical density. However, when these masses are scaled to $M_{200}$ (approximately 16\% lower than $M_{\rm vir, tot}$), these total mass estimates are in excellent agreement with our results, where $M_{200, \rm tot} = 1.0-1.3 \times 10^{12}M_\odot$.
  
\cite{callingham19} use satellite kinematics to measure the Milky Way mass, thus, as the satellites extend out to the virial radius,  their measure is a direct measure of the total mass. Their derived total mass and dark halo concentration, $M_{200, \rm tot} = 1.17^{+0.21}_{-0.15}$, $c_{200} = 10.9^{+2.6}_{-2.0}$, are in good agreement with our results. This agreement is particularly pleasing as the authors quote one of the most precise and accurate total mass measurement to-date, and use a completely different analysis technique (and dynamical tracers) to derive the mass.

These results imply that we are generally converging to a total Milky Way mass of $M_{200, \rm tot} \sim 1 \times 10^{12} M_\odot$. This mass, which is on the low end of the wide spectrum of advocated masses, effectively bails the Milky Way out from the ``too big too fail" problem.  \cite{purcell12} and \cite{wang12} showed that the number of massive satellites predicted around $\sim 10^{12}M_\odot$ haloes is in good agreement with the Milky Way dwarf population. In contrast, many more massive subhaloes are predicted to reside in more massive host haloes, which led to the original conundrum posed by \cite{boylan12}. Our total Milky Way mass also has implications for the identity of the dark matter \citep[e.g.][]{kennedy14, lovell14}, the influence of reionization on the dwarf satellite population \citep{bose18}, and the uniqueness of some of the satellite dwarf galaxies (e.g. the Magellanic clouds and Leo I, \citealt{boylan13, cautun14}). Indeed, the wide-range of Milky Way mass estimates quoted in the literature has allowed this parameter to frustrate our investigations into apparent small scale problems with the $\Lambda$CDM model; now in the era of \gaia\ we can hope to remove, or at least narrow down, this important degree of freedom in future analyses.

\section{Conclusions}
\label{sec:conc}

In this work, we have investigated the high velocity tail of local Galactic halo stars using a combination of analytical models, cosmological simulations and 6 dimensional \gaia\ data. We make use of recent constraints on the origin of the inner stellar halo, which affects the velocity distribution of the halo stars, to construct a prior on the shape of the high velocity tail. We use this insight to estimate the local Galactic escape speed, and relate this measurement to the total Milky Way mass. Our main conclusions are summarised as follows:

\begin{itemize}

\item Using simple, analytical models we show that the shape of the high velocity tail is strongly dependent on the velocity anisotropy and density profile of the halo stars. We find that for a fixed gravitational potential, systems with highly radial velocity anisotropy and/or shallow density profiles have more extended velocity tails.

\item The shape of the high velocity tails in the Auriga simulations agree with the predictions from the analytical models. We further find that the assembly history of the halo, namely the mass and epoch of the most massive dwarf satellite mergers, impacts the form of the high velocity tail. We also use the simulations to define the outer radial boundary for the escape velocity. An appropriate choice, based on the orbits of the stars in the simulations, is $2r_{200}$.

\item By modeling the high velocity tail with a functional form $\propto \left(v_{\rm esc}-v\right)^k$ \citep{lt90}, we use the simulations to construct an appropriate prior on $k$. Recent observations of highly eccentric orbits in the inner halo, caused by a massive, early accretion event,  conspire to form a prior appropriate for relatively extended velocity tails, with $1 < k < 2.5$. This allowed range of $k$ is lower than previous priors derived from cosmological simulations \citep{smith07, piffl14}, as these works consider the entire range of assembly histories available rather than the particular case of the Milky Way.

\item We apply our formalism to \gaia\ DR2 and measure a local escape velocity of $v_{\rm esc}(r_0) = 528^{+24}_{-25} \kms$. We use the definition of the escape boundary ($2r_{200}$) to relate this measurement to the total Milky Way mass.  By combining our escape velocity measurement with the local circular velocity ($v_c(r_0) = 230 \kms$, \citealt{eilers18} ), we find $M_{200, \rm tot} = 1.00^{+0.31}_{-0.24} \times 10^{12}M_\odot$, and $c_{200} = 10.9^{+4.4}_{-3.3}$. Our mass and concentration measurements are in good agreement with \cite{callingham19} (see also \citealt{patel18}), who use a completely independent methodology to model the dynamics of satellite galaxies out to the virial radius of the Galaxy.

\end{itemize}

\noindent
The premise of this work is to use our knowledge of the assembly history of the Milky Way halo, and the corresponding phase-space distribution of halo stars, to inform our modeling of the high velocity tail, and hence place a stronger constraint on the mass of the Milky Way. In the past months since the first astrometric \gaia\ data release, our knowledge of the Milky Way halo has increased dramatically. Now we can start to use that knowledge to inform our models, and reduce the wide parameter space set by cosmic variance. In the present application to the high velocity tail, the Universe has conspired to be kind to us. The dominance of an early, massive accretion event, and the resulting highly eccentric orbits of the halo stars leads to an extended, and well-defined high velocity tail. This fortuitous situation allows us to make a robust measurement of the local escape velocity, and hence the total Milky Way mass. The future \gaia\ data releases will continue to further our knowledge, and place even tighter constraints on these fundamental parameters.

\section*{Acknowledgements}
We thank the referee of our paper, Matthias Steinmetz, for his insightful comments and suggestions.

AD is supported by a Royal Society University Research Fellowship.  AF is supported by a European Union COFUND/Durham Junior Research Fellowship (under EU grant agreement no. 609412). AD and AF also
acknowledge the support from the STFC grant ST/P000541/1. The research leading to these
results has received funding from the European Research Council under
the European Union's Seventh Framework Programme (FP/2007-2013) / ERC
Grant Agreement n. 308024.

This work has made use of data from the European Space Agency (ESA) mission
\gaia\ (\url{https://www.cosmos.esa.int/gaia}), processed by the \gaia\
Data Processing and Analysis Consortium (DPAC,
\url{https://www.cosmos.esa.int/web/gaia/dpac/consortium}). Funding for the DPAC
has been provided by national institutions, in particular the institutions
participating in the \gaia\ Multilateral Agreement.

This work
used the DiRAC Data Centric system at Durham University, operated by
ICC on behalf of the STFC DiRAC HPC Facility (www.dirac.ac.uk). This
equipment was funded by BIS National E-infrastructure capital grant
ST/K00042X/1, STFC capital grant ST/H008519/1, and STFC DiRAC
Operations grant ST/K003267/1 and Durham University. DiRAC is part of
the National E-Infrastructure.

AD thanks the staff at the Durham University Day Nursery who play a key role in enabling research like this to happen.

\bibliographystyle{mnras}
\bibliography{mybib}

\bsp	
\label{lastpage}
\end{document}